# Compositional Diversity of Rocky Exoplanets

Keith Putirka[1], Caroline Dorn[2], Natalie Hinkel[3], Cayman Unterborn[4]

[1] Department of Earth and Environmental Sciences, California State University, Fresno, 2576 E. San Ramon Ave, MS/ST 24, 93740
[2] University of Zurich, Institute of Computational Sciences, Y11-F74, Winterthurerstrasse 190 ch-8057, Zurich
[3] Southwest Research Institute, 6225 Culebra Rd, San Antonio, Texas 78228-0510
[4] School of Earth and Space Exploration, Arizona State University, PO Box 876004, Tempe, AZ, 85287-6004

**ABSTRACT**

Star compositions are essential for examining densities and compositional ranges of rocky exoplanets, testing their similarity to Earth. Stellar elemental abundances and planetary orbital data show that of the ~5000 known minerals, exoplanetary silicate mantles will contain mostly olivine, orthopyroxene, and clinopyroxene, ± quartz, and magnesiuwüstite at the extremes; wholly exotic mineralogies are likely absent. Understanding these exotic geological systems requires a better marriage of geological insights to astronomical data. The study of exoplanets is like a mirror, reflecting our incomplete understanding of Earth and neighboring planets; new geological/planetary experiments, informed by exoplanet studies, are needed for effectual progress.

## 1. Introduction

The disciplines of geology and astronomy have existed at a considerable remove for >300 years. The discovery of rocky exoplanets during the last two decades, however, requires the closing of that distance: knowledge and methods in one discipline can affect hypotheses in the other. The limitations in each field drive innovation in the other. No better example is the interpretation of rocky exoplanet compositions, the range and geologic and geophysical consequences of which we examine in several respects.

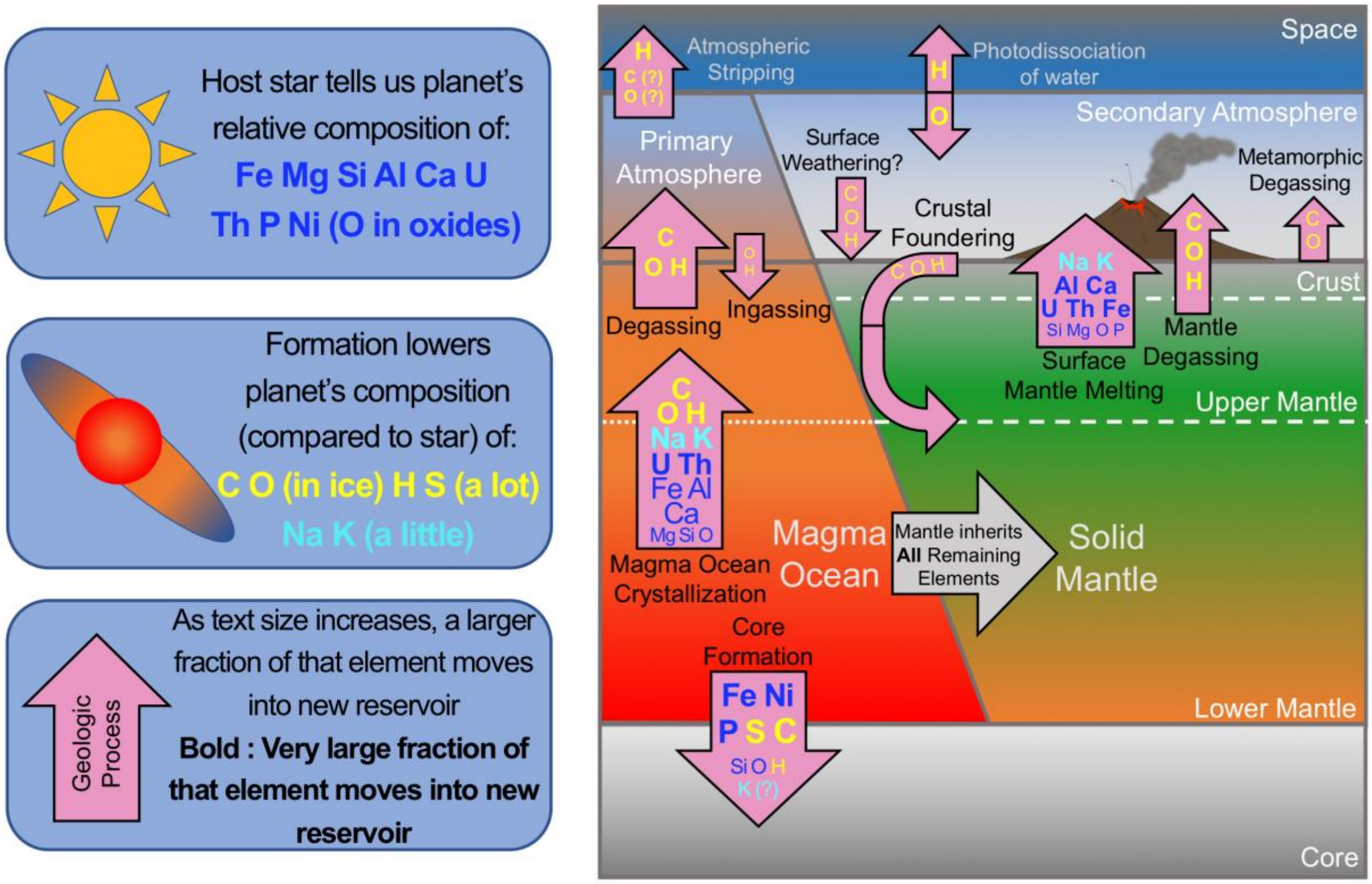

Figure 1. Cartoon of all geochemical and geophysical processes that partition elements between the different major reservoirs for a rocky exoplanet in both the hotter, magma ocean (reddish part of right panel) and cooler, solid-mantle (greenish part of right panel) phases of planetary evolution.

## 2. Planetary Building Blocks: the Range of Elemental Abundances Among Nearby Stars

Although we can measure the makeup of exoplanetary atmospheres, the compositions of planetary surfaces and interiors are inaccessible. And yet, understanding the make-up of a planet is incredibly important since the overall geologic-state of a rocky planet is a function of its bulk composition, internal structure, mantle and crust mineralogy, and the makeup of its atmosphere (e.g., Foley & Smye, 2018). In the same way that the Sun and the Solar System planets all formed from the same protoplanetary disk (Dorn et al. 2015; Unterborn et al. 2016), stars and exoplanets are chemically intertwined (e.g., Bond et al. 2010; Thiabaud et al. 2015) because they are born from the same cloud of interstellar dust (Figure 1). Therefore, we make a fundamental assumption that rocky planet compositions are similar to the stars they orbit. Unlike the volatile elements (e.g., C, N, O), the major (Mg, Si, Fe) and minor (Al, Ca) rocky-planet element abundances vary by less than a factor of three between the Sun, Earth and Mars. And compared to the volatile elements, refractory components are not expected to considerably fractionate *relative to each other during formation*. Mercury, with it's Fe-rich composition does not fit this trend, and the reason why is an active area of research. So, while the Sun has orders-of-magnitude more atoms of Mg than Earth, the molar Fe/Mg of Earth and Sun are quite close (0.9 versus 0.83, respectively).

To that end, the *range* of elemental abundances observed in nearby stars has a direct impact on the variations that can be expected for interior composition of orbiting rocky planets. For stars, elements are measured as the number of atoms per unit volume (number density), indicated as $N_{Fe}$ for e.g., Fe. Abundances are almost always expressed as a ratio of one element to another, so for a star, the parameter $(N_{Fe}/N_H)_{Star}$ shows how the amount of Fe in a star varies with respect to H, which is considered to be relatively constant in the Universe with respect to time. This ratio is then compared to the same ratio in the Sun's photosphere, as a normalization, that is expressed using the dex (decimal exponent) system indicated by the square brackets: [Fe/H] = $\log(N_{Fe}/N_H)_{Star} - \log(N_{Fe}/N_H)_{Sun}$. Examining the range of these abundances, we look to the Hypatia Catalog, a multi-dimensional database of stellar elemental abundances that currently spans 78 elements in >9400 stars within 500 pc (or ~1600 lightyears) of the Sun (from +215 literature sources; see Hinkel et al. 2014 and www.hypatiacatalog.com). Given the breadth of this database, the median value is used when multiple groups measure the same element within the same star (Hinkel et al. 2016) and all measurements are normalized to the same solar composition, e.g., Lodders 2009. The Hypatia Catalog, as a wholly self-consistent database, allows us to  estimate the range of potential exoplanet compositions.

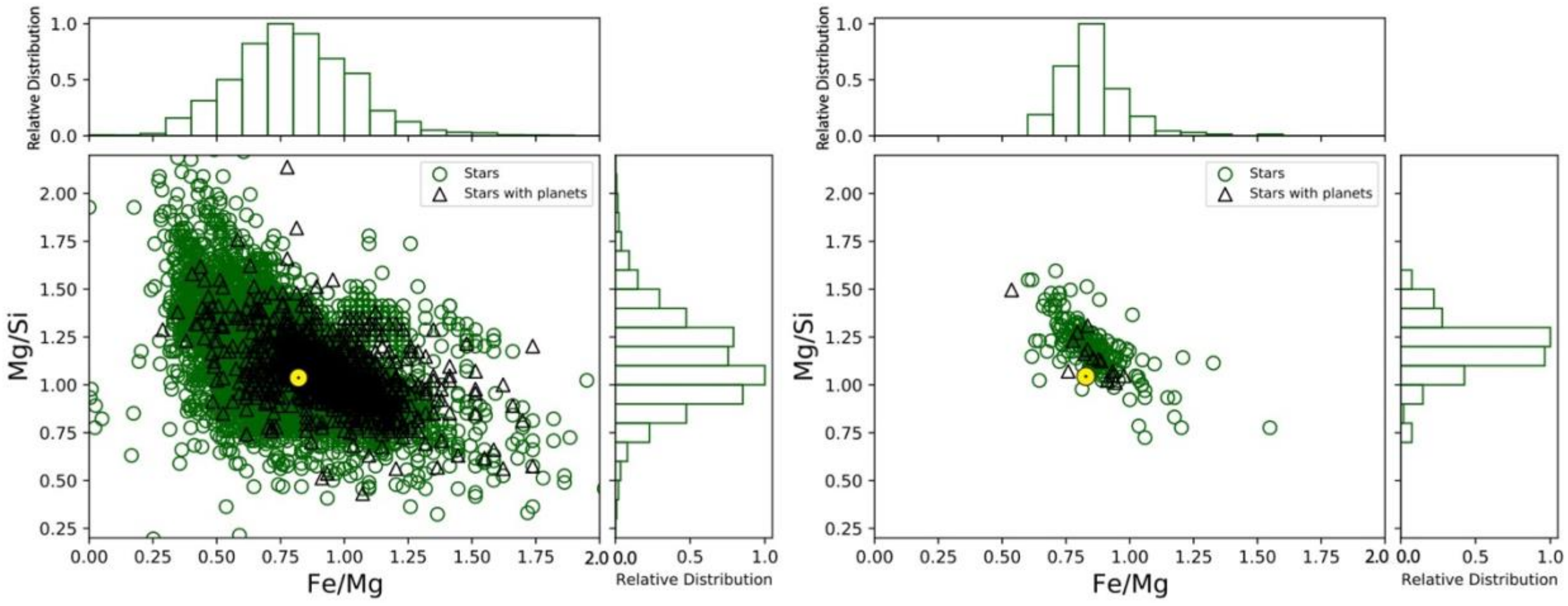

Figure 2 – Molar ratios Fe/Mg vs Mg/Si for full Hypatia Catalog sample (A, Hinkel et al. 2014) while the right column (B) gives only the solar twins (stars where effective temperature, surface gravity, and [Fe/H], are very similar to the Sun; (Fe/Mg)) ⊙ = 0.832; (Mg/Si) ⊙ = 1.023; yellow dot]. Despite ~~the fact~~ the similarity between solar twins and the Sun, there is still quite a lot of intrinsic variation between the stars. Exoplanet host stars are indicated with triangles and also show divergence in their molar ratios.

Elements that play the largest role in determining the internal structure and mineralogy of a rocky planet, e.g., Mg, Si, and Fe, have dex values (e.g., [Mg/H]) that range more than four orders of magnitude within the full Hypatia Catalog, from 1.0 to -3.0 dex (10-1 – $10^{-3}$) and below. As molar fractions, the majority of stars in the Hypatia Catalog fall within Mg/Si = 0.7–1.5, and Fe/Mg = 0.4–1. (see Figure 2A). This large variation in molar fractions also exists for those stars that host exoplanets. However, the range of the molar fractions seen in Figure 2A may be the result of 1) complex stellar physics, which affect the elements within the photosphere, since the Hypatia Catalog contains many stellar types that span a range of ages, masses, and temperatures; and 2) influences from galactic chemical evolution since stars could have originated from different parts of the galactic disk with varying compositions and evolutions.

To simplify matters, we can focus on solar twins: stars within 100 pc (or ~325 lightyears) that have similar temperatures and surface gravities to the Sun and an [Fe/H] within 0.1 dex of solar (per Bedell et al. 2018). The solar twins in the Hypatia Catalog, despite properties similar to the Sun, still show a wide range of stellar elemental abundances: [Mg/H] = [-0.16, 0.24], and [Si/H] = [-0.16, 0.11] dex, or roughly 0.7 to 1.5 times more atoms of these elements than the Sun. Even individual stars have a mixture of super- and sub-solar stellar abundances for these elements. As seen in Figure 2B, the solar twins' molar fractions cover a much smaller range of Mg/Si and Fe/Mg compared to the full catalog. But there are a handful of stars with large molar Fe/Mg (> 1.2), some of which are a result of the mixed super- and sub-solar abundances. It is clear then, that even when analyzing stars with very similar properties to one another, there is intrinsic variation in the abundances of those elements that primarily go into building rocky exoplanets. This variation will directly impact the bulk composition and structure (Section 3) and dynamical behavior (Section 4) of their orbiting exoplanets.

## 3. Inferring Exoplanet Compositions From Mass-Radius Relationships

In addition to using star compositions, we can infer a rocky exoplanet's composition by measuring planetary mass and radius. We can obtain radius from the so-called 'transit method', when an exoplanet passes in front of its star, causing a partial extra-solar eclipse; mass is obtained by the 'radial velocity method' which uses shifts in a star's position due to gravitational tugs by an orbiting planet. In rare cases, both can be measured.

The majority of planets discovered in our stellar neighborhood are larger than Earth in both mass and radius ('super-Earths'), yet smaller than Neptune ('mini-Neptunes'). While mini-Neptunes have thick H/He atmospheres, super-Earths may have lost them within the first ~100 Myr of their evolution and are thus limited in size with radii smaller than 2 Earth-radii ($R_\oplus$, Fulton et al. 2017). The diversity among super-Earth and mini-Neptune interiors is largely unknown. The challenge in determining planetary interiors originates from not only limited observables for an individual exoplanet, but also from the typically large uncertainties in the measured properties.

Super-Earths where both masses and radii are known (Fig. 3) show densities that are both higher and lower than the density of Earth (red solid line). This diversity must reflect relative amounts of Fe, Mg and Si, as well as volatile contents (e.g., $H_2O$). The majority of planet-hosting stars have molar Mg/Si between 0.7—1.5, and molar Fe/Mg between 0.4--1.5 (Fig. 2). Translated into a mass and radius diagram (Fig. 3), these stellar ranges imply a variability of 25% in planet mass for a given radius, or 7% in radius for a given mass, provided that super-Earths have pure-Fe cores and Fe-free silicate mantles.

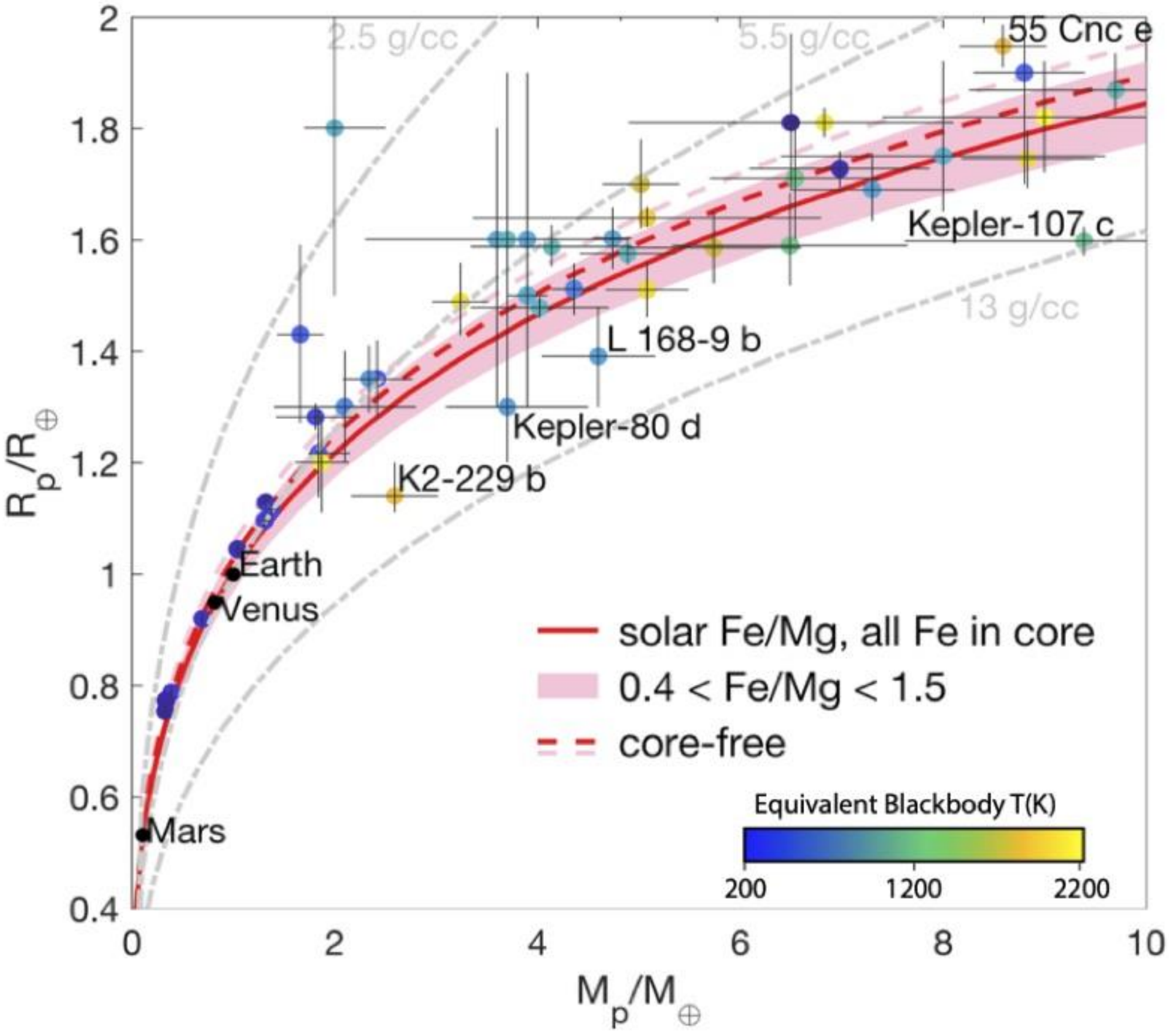

Figure 3. Mass ($M_p$) and radius ($R_p$) of confirmed exoplanets from Otegi et al. (2020) in units relative to the mass ($M_\oplus$) and radius ($R_\oplus$) of Earth. We show the mass-radius curves corresponding to the range of Fe/Mg as taken from planet-hosting stars (Hinkel et al. 2014). Earth’s mass and radius is denoted by $M_\oplus$ and $R_\oplus$, respectively. Color coding indicates planetary blackbody temperatures (zero albedo, and no greenhouse gases or internal energy sources) based on orbital distance and stellar temperature.

Uncertainties of stellar elemental abundances are assumed to be zero. Iron is assumed to be in the core (red solid lines) or fully oxidized in the mantle (red dashed lines). Only those planet names are given that are discussed in the text. In light grey, three iso-density curves are shown.

The super-Earths of Fig. 3 show a wider distribution of densities than the bulk densities inferred from stellar elemental abundances alone. Although only a few of these super-Earths have host-star element abundances that have been determined, the range of stellar Fe/Si and Fe/Mg (e.g., Fig. 1) are smaller than that inferred from Super-Earth masses and radii. Nevertheless, comparing those few systems where available host-star elemental abundances are available, Schulze et al. (pers. communication) find a direct correlation between planet densities and stellar compositions, with one exception: K2-229b. But additional high-quality data are needed to test the extent to which star and planet compositions are correlated.

Mass and radius also allow us to calculate a planet's mean density and infer its interior structure via the "mass-radius relationship" (e.g., Jontoff-Hutter, 2019). However, mean density is degenerate in that it does not distinguish between different planetary components, such as the metallic core, silicate crust + mantle, and volatile layers (e.g., Dorn et al., 2015; Unterborn et al., 2018). However, even if stellar elemental abundances are available, to adopt as a proxy for a planet's bulk composition, we are still left with a 'degenerate inference' on the planet's structure. For example, iron can be in the core (as Fe), or oxidized in the mantle (as FeO), which translates to a difference in calculated density of approximately 10% (Fig. 3, dashed versus solid red lines). This balance of Fe depends upon the availability of oxygen, and is often expressed as oxygen fugacity, or $fO_2$, which for exoplanets may be roughly similar to rocky worlds in our Solar System (Doyle et al. 2019). Whether a planet is solidified or retains a magma ocean can also lead to density variations of up to 13 % (Bower et al. 2019). There are good candidates for magma ocean planets among the highly irradiated super-Earths (those with unusually low densities, e.g., 55 Cnc e in Fig. 3), since thick volatile envelopes are unlikely to have survived on these hot worlds which could otherwise explain their low bulk densities. Other volatile-poor worlds that have been hypothesized to exist are carbon-rich planets (Madhusudhan et al. 2012) and Ca-Al-rich planets (Dorn et al. 2019) where bulk densities can be significantly lower than Earth, by 10-20%.

Even planets with major and minor element compositions matching their host star’s can still be relatively low in density if they are volatile-rich (e.g., $H_2O$, $CO_2$, $CH_4$). The radii of such planets are sensitive to both the amounts of volatiles and the volatile species, as well as planetary thermal states and stellar irradiation. For this reason, the distribution of planet densities extends to lower values compared to the density range inferred from stellar elemental abundances. For example, the small addition of 0.01 wt% of a steam ($H_2O$) atmosphere can increase a radius by ~10 %. However, among the diverse atmospheres of the terrestrial planets of the Solar System, there is no planet whose volatile envelope contributes more than 2% to its radius. The diversity of water budgets and atmospheres on Super-Earths and terrestrial-sized planets is a matter of debate, however rocky planets forming significant volatile fractions seem likely.

For the most dense planets in Fig. 3, it remains to be seen if or why their Fe/Mg and Fe/Si ratios depart from their host star counterparts. Planets that are highly depleted in Mg and Si and enriched in Fe might be difficult to form, even when accounting for giant impacts that could strip them of their silicate mantles. A giant impact may have caused Mercury’s high iron content by ablating it’s silicate mantle, but details of this model have yet to be explained. And giant impacts are a less probable cause as planet size increases. Understanding the Fe-enrichment of the planets Kepler-107c, L1689b, Kepler-80d, and K2-229 b will be crucial to further understand how planet formation can shift a rocky planet’s composition away from that of its host star. However, while stellar composition may not serve as a direct guidepost to a particular planet’s bulk composition, their ranges appear to inform us of the *range* of rocky exoplanet compositions.

## 4. Compositional Consequences: Implications for Tectonics and Continental Crust

The question of whether or not a given planet might exhibit plate tectonics depends upon the material properties of a planet’s interior, which control mantle convection and partial melting, and the kinds of planetary crusts, atmospheres and oceans that can be formed. These material properties depend upon the kinds of minerals—and elements—that comprise a silicate mantle, and so there is a direct link between planetary habitability and a planet’s bulk composition as inherited from its host star. As one might expect, nucleosynthesis provides only so many ingredients from which a planet might form. The twenty-one most abundant elements in the Sun are, in order: H, He, O, C, Ne, N, Mg, Si, Fe, S, Ar, Al, Ca, Na, Ni, Cr, Mn, P, Cl, K and Ti (Lodders 2009); ignoring the volatile H, He, C, Ne, N and S (which will be depleted in rocky planets; see Putirka and Rarick (2019) for the cases of C and S) all of O, Mg, Si and Fe comprise 88% of the remainder. These four elements, with varying amounts of the remainder, are the dominant planetary ingredients.

Iron is of special interest as it occurs in both the core, as Fe metal ($Fe^{o}$), and in the mantle, as FeO ($Fe^{2+}$), which then can combine with $SiO_2$, MgO and CaO to form olivine $(Mg,Fe)SiO_4$, orthopyroxene $(Mg,Fe)_2SiO_6$, clinopyroxene, $Ca(Mg,Fe)Si_2O_6$, and magnesowüstite (Mg,Fe)O. For a given bulk Fe content, the more Fe that occurs as FeO, the smaller the core and the less dense the planet (see Section 2). This partitioning of Fe is related to $fO_2$ and early exoplanetary atmospheres (Schaefer and Fegley, 2017). Because exoplanetary $fO_2$ is similar to our own solar system (Doyle et al. 2019), we can use Mercury and Mars as proxies. For example, Mercury’s mantle appears to have almost no FeO (Nittler et al. 2018), whereas Mars’s mantle is FeO-rich. Our proxy for $fO_2$, then, is the ratio $\alpha_{Fe} = \frac{Fe^{mantle}}{Fe^{whole\ planet}}$, i.e., the fraction of elemental Fe in a silicate mantle relative to total planetary Fe, which is 0 for Mercury, 0.54 for Mars and ~0.27 for Earth. Figure 4A shows the compositions of rocky mantles when bulk compositions are solar and $\alpha_{Fe}$ varies from 0.0 to 0.6. Having no FeO in its mantle, Mercury’s case is degenerate: it is highly enriched in Fe relative to Solar due to its large core. However, Earth, Mercury, Mars and Moon all have silicate mantles that fall encouragingly close to the Solar trend—about where expected, given the sizes of their metallic cores. Earth is furthest from the trend, falling to the $SiO_2$-poor side of the Sun and chondrites (all calculated at $\alpha_{Fe}$=0.27), even though we add 7% Si to the core (see Putirka and Rarick 2019), and no hidden component can make up the difference (Putirka and Rarick 2019).

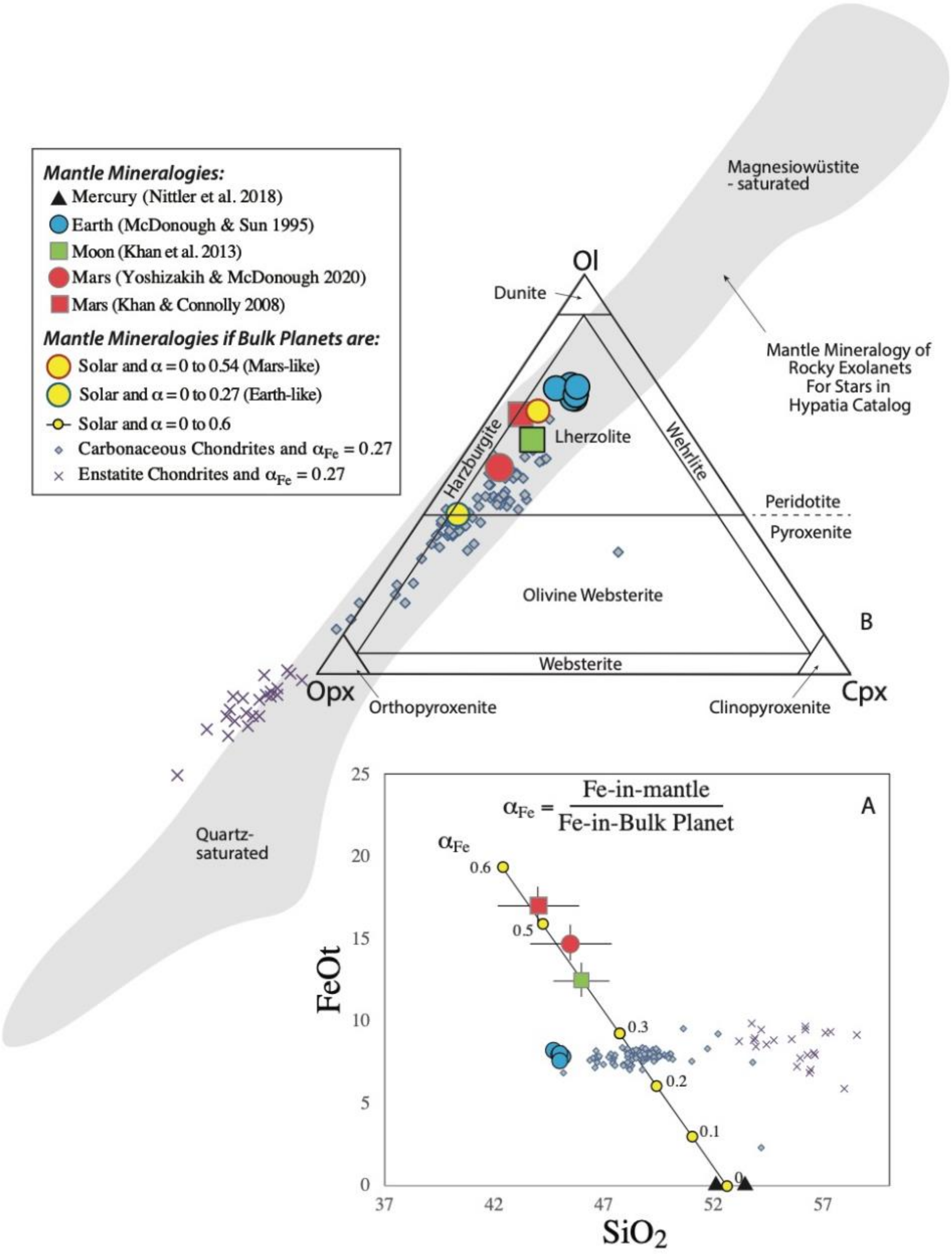
Mantle Mineralogies:
Mercury (Nittler et al. 2018)
Earth (McDonough & Sun 1995)
Moon (Khan et al. 2013)
Mars (Yoshizakih & McDonough 2020)
Mars (Khan & Connolly 2008)
Mantle Mineralogies if Bulk Planets are:
Solar and α = 0 to 0.54 (Mars-like)
Solar and α = 0 to 0.27 (Earth-like)
Solar and α = 0 to 0.6
Carbonaceous Chondrites and αFe = 0.27
Enstatite Chondrites and αFe = 0.27
Magnesiowüstite - saturated
Mantle Mineralogy of Rocky Exolanets For Stars in Hypatia Catalog
Ol
Dunite
Harzburgite
Lherzolite
Wehrlite
Peridotite
Pyroxenite
Olivine Websterite
Websterite
B
Opx
Cpx
Orthopyroxenite
Clinopyroxenite
Quartz-saturated
αFe = Fe-in-mantle / Fe-in-Bulk Planet
A
FeOt
SiO2
0.6
0.5
0.3
0.2
0.1
0
25
20
15
10
5
0
37
42
47
52
57

Figure 4. A: Weight % $SiO_2$ vs. FeOt (total Fe as FeO in a silicate mantle). Solar photosphere elemental abundances are converted into oxides (yellow circles; see Putirka and Rarick 2019), after allowing some Fe to enter the core ($\alpha_{Fe}$ varies from 0.0 to 0.6); mantle compositions from chondritic meteorites use an Earth-like $\alpha_{Fe}$ = 0.27. Solar-derived mantle compositions (yellow circles) are compared to bulk silicate Earth (McDonough and Sun 1995), Mercury (Nittler et al. 2018) and Mars (Khan and Connolly 2008; Yoshizaki and McDonough 2020). Note: bulk Mercury has excess Fe relative to Solar due to its large core; Earth has low $SiO_2$ despite allowing 7% Si in the core. B: oxides are converted to mineral abundances olivine (Ol), orthopyroxene, (Opx) and clinopyroxene (Cpx) (see Putirka and Rarick 2019). All data have mineral proportions that sum to 1; within the triangle, amounts of Ol, Cpx and Opx are positive; outside the triangle, data have negative amounts of either Ol or Opx, which can be equated to positive amounts of quartz or magnesiowüstite respectively. Gray field shows exoplanetary range when $\alpha_{Fe}$ = 0.27. Note that small differences in major oxides (Earth vs. Sun at $\alpha$ = 0.27; 3A) translate to large differences in mineralogy.

Figure 4A thus lends confidence to using star compositions from the Hypatia Catalog to estimate (theoretical) rocky planet mineralogies, assuming that their non-volatile element abundances mimic the stars they orbit. High abundances of Mg, Si, and Fe (e.g., Fig. 4A) translate to mantles dominated by Opx and Ol (Fig. 4B). Calcium and Al are sufficiently abundant to allow for small amounts of clinopyroxene (Cpx) and garnet $(Mg,Fe)_3Al_2Si_3O_{12}$. But Na, Al, Ti, Cr, etc. are too low in abundance to yield utterly exotic mineralogies (no exoplanets are made of corundum, $Al_2O_3$, rutile, $TiO_2$, chromite, $FeCr_2O_4$ or nepheline, $NaAlSiO_4$, etc.). Indeed, only two parameters, $\alpha_{Fe}$ and the ratio $(FeO+MgO)/SiO_2$, are needed to define mantle mineralogy (Putirka and Rarick 2019), and exoplanetary mantle rock types will mostly range from orthopyroxenite to dunite. Some are sufficiently $SiO_2$-rich to contain quartz ($SiO_2$, Qz) in their upper mantles, while others are sufficiently FeO- and MgO-rich to contain magnesiowüstite (Fig. 4B), but even these cases are rare.

While the mineralogical variety (Fig. 4B) is seemingly narrow, there may be curious geological consequences. Weller & Lenardic (2018) show that "yield stress" (the amount of stress required to deform a mineral) controls whether a planet exhibits a "mobile lid" (plate tectonics, like Earth) or a stagnant lid (like Earth's Moon or Mars) regime. And because some experimental data indicate that Opx is stronger than Ol at high temperatures, plate tectonics might be inhibited nearer the Opx apex (Fig. 4A) [see Ballmer and Noack, this issue]. But since Qz is very weak, even small amounts might facilitate plate tectonics for compositions below the Opx apex.

However, crust formation might confound this view: Opx- and Qz-rich mineralogies melt at lower temperatures than Ol-rich compositions (e.g., Lambart et al. 2016) and may yield thicker crusts. A thick crust of any type could induce density and stress contrasts that enable tectonic activity, despite a strong lithosphere. Relevant partial melting and stress-strain experiments are almost nonexistent, but are needed to delimit exoplanetary crust and mantle dynamics, which will in turn govern near-surface abundances of life-supporting elements: C, N, P, and K.

Of final but critical concern are the concentrations of heat producing elements, and their impact on the Urey ratio (Ur; the ratio of internal heat production to total surface heat flux). Higher concentrations of U, Th and K mean a higher Ur, which results in a hotter planetary interior that is more able to convect; a higher Ur might support plate tectonics, which would affect global water and C cycles [see Ballmer and Noack, this issue]. Unterborn et al. (2015) made the first study of Th concentrations using a sample of “Solar twins. Thorium is a useful proxy for U and K as all three have similar geochemical behaviors, except that K is more volatile during accretion. Unterborn et al. (2015) find that solar twins have Th contents ranging from 59% to 251% of Solar. As Th is a refractory element, we suspect that their orbiting planets should exhibit the same, and that Ur ratios vary proportionally. The two-fold challenge is to determine whether Ur and mantle mineralogy vary independently, and to infer how convective and tectonic behaviors would be affected.

## 5. Filling the gaps to define “Earth-like” Planets

As we probe atmospheric compositions for potentially “Earth-like” worlds with the space-based instruments, like the James Webb Space Telescope (JWST), as well as many new ground-based telescopes in the coming decades, we will see an exoplanet in only a single snapshot in time. It remains, then, for geoscientists to use this datapoint and the other scant data available for that system (mass, radius, host-star age, etc.), to tell the story of the exoplanet’s history and how it reached that point in its evolutionary path.

It is becoming increasingly clear that compositional diversity will play a role in determining whether a rocky exoplanet is indeed Earth-like in the “geologic” sense. In other words, rocky planet composition will determine whether it evolved and behaves like the Earth, as opposed to merely having the same mass, radius, or core mass fraction. Planetary evolutionary history, however, is quite complex. While the bulk composition of a planet sets its mineralogy and volatile content, a myriad of processes move each of the constituent elements within the planet between the core, mantle, crust, and atmosphere. No element is unique, as some tend to prefer the core (e.g., P, Fe, Ni,), while others quickly migrate to the atmosphere (Figure 1).

A rocky exoplanet’s geologic story begins during its magma ocean phase, assuming it is Moon-size or larger. During this early phase of planetary evolution the central metallic core forms, which can permanently sequester many elements. Even biocritical elements such as C, O, and P enter into the core, which could yield a planetary surface less conducive to life (e.g., Stewart & Schmidt, 2007; Fischer et al., 2020). As a magma ocean solidifies, elements that are incompatible in both the metallic core and mantle silicate minerals are partitioned into the atmosphere, where stellar winds may strip them from the planet entirely. For planets that are Earth-sized or larger, the residual “depleted mantle” will continue to evolve both chemically and physically via convective forces, leading to continued volcanism and degassing that shifts elements from the mantle to the basaltic oceanic crust (and potentially into continental crust, however, the production of andesitic crusts is not well understood for exoplanets). If tectonic processes become active (e.g., Weller and Lenardic 2018; Foley & Smye, 2018), physical and chemical weathering will partition volatiles from the atmosphere back into surface rocks that would then be transported back into the deep mantle via crustal foundering either through stagnant lid or plate tectonics. From all of this, an atmosphere is born and evolves. To understand this planet, and indeed whether it is at all “Earth-like,” we must tell this story using only a single coarse measurement of atmospheric composition at a single point in time. Difficult? Yes. Impossible? Perhaps not.

When beginning to explore these complex planetary processes, we must understand that vastly more is known about the nature and time dependence for the Earth and its immediate rocky neighbors (perhaps with the exception of Venus), which has been acquired through hundreds of years of planetary science and geological research. Applying these insights to those planets of near- and non-Earth composition is difficult due to the dearth of data necessary to model their geologic history, such as melting curves, thermoelastic mineral properties and volatile solubilities for non-Earth-like bulk compositions (Fig. 3). Morbidelli et al. (2016; their Fig. 3B) show that accretionary processes play a major role in determining a planet's bulk composition and volatile budget. How accretion and related processes affect the amounts of water, carbon and other geologically important elements, such as Na and K, that are delivered to rocky exoplanets is an active area of research. The sheer diversity of rocky exoplanets in both size and composition are showing us that there are considerable gaps in our underlying datasets, but also how our comparatively robust knowledge of the Earth and our Solar System can limit, or even bias, our intuition for conceptualizing rocky planet evolution *in general.* The study of rocky exoplanets frees us from parts-per-million level precision, but at the cost of large uncertainties and little directly observed data.

Some nearby frontiers that can dramatically improve our understanding of exoplanets. For example, the most exciting developments in exoplanetary studies are likely to come from investigating our very non-Earth-like neighbor, Venus. Its tectonic state and geologic history are likely quite different, but we are almost utterly ignorant of its bulk composition, the age and composition of its crust and surface, the size of its core, and the nature of mantle convection. We also need new experiments, to determine the phase diagrams and rock strengths for bulk compositions that are not Earth-like, but relevant to many exoplanets. Geologists and astronomers will need to work together to collect and interpret data that will have the greatest and widest impact for understanding planetary systems, in this and other stellar systems.

The story of rocky exoplanets and the meaning of their diverse compositions cannot be penned by any one author, but will must be created by a unified field of exoplanetary science that involves both geologists and astronomers.

## Acknowledgments

Putirka acknowledges support from NSF grants -1921182, -1250323 and -1250322.

## References Cited

1. Bedell, M, Bean, JL, Melendez, J, Spina, L, et al. (2018) The Chemical Homogeneity of Sun-like Stars in the Solar Neighborhood. Astrophysical Journal 865:68-81
2. Bond, JC, O'Brien, DP, Lauretta, DS (2010) The Compositional Diversity of Extrasolar Terrestrial Planets. I. In Situ Simulations. The Astrophysical Journal 715: 1050-1070

3. Bower D.J, Kitzmann D, Wolf AS, Sanan P, Dorn C, Oza AV (2019) Linking the evolution of terrestrial interiors and an early outgassed atmosphere to astrophysical observations. Astronomy & Astrophysics, 631, A103
4. Dorn C, Harrison JH, Bonsor A, Hands TO (2019) A new class of Super-Earths formed from high-temperature condensates: HD219134 b, 55 Cnc e, WASP-47 e. Monthly Notices of the Royal Astronomical Society, 484, 712
5. Dorn C, Khan A, Heng K, Connolly JA, Alibert Y, Benz W, Tackley P (2015) Can we constrain the interior structure of rocky exoplanets from mass and radius measurements? Astronomy & Astrophysics, 577, A83, 1-18
6. Doyle AE, Young ED, Klein B, Zuckerman B, Schlichting HE (2019) Oxygen fugacities of extrasolar rocks: evidence for Earth-like geochemistry of exoplanets. Science 366:356-359.
7. Fischer RA, Cottrell E, Hauri E, Lee KKM, Le Voyer M (2020) The carbon content of Earth and its core. Proceedings of the National Academy of Sciences 117:8743-8749
8. Foley, B. J., & Smye, A. J. (2018) Carbon Cycling and Habitability of Earth-Sized Stagnant Lid Planets. Astrobiology 18:873-896
9. Fulton BJ, Petigura EA, Howard AW, et al. (2018) The California-Kepler Survey. III. A Gap in the Radius Distribution of Small Planets. The Astronomical Journal 154:109-138
10. Hinkel NR, Timmes F, Young PA, Pagano MD, et al. (2014) The Chemical Homogeneity of Sun-like Stars in the Solar Neighborhood. The Astronomical Journal 148:54-67
11. Hinkel NR, Young P.A, Pagano MD, et al. (2016) A Comparison of Stellar Elemental Abundance Techniques and Measurements. The Astrophysical Journal Supplemental 226:4-70
12. Jontoff-Hutter J. (2019) The Compositional Diversity of Low-Mass Exoplanets. Annual Review of Earth and Planetary Sciences 47:141-171
13. Khan A, Connolly JAD (2008) Constraining the composition and thermal state of Mars from inversion of geophysical data. Journal of Geophysical Research, Planets 113:E07003
14. Khan A, Pommier A, Neumann GA, and Mosegaard K (2013) The lunar moho and the internal structure of the Moon: a geophysical perspective. Tectonophysics 609:331-352
15. Lambart S, Baker MB, Stolper EM (2016) The role of pyroxenite in basalt genesis: melt-PX, a melting parameterization for mantle pyroxenites between 0.9 and 5 GPa. Journal of Geophysical Research 10.1002/2015JB012762
16. Lodders, K (2009) Solar system abundances and condensation temperatures of the elements. The Astrophysical Journal 591:1220-1247
17. Madhusudhan N, Lee KK, and Mousis O (2012) A Possible Carbon-rich Interior in Super-Earth 55 Cancri e. The Astrophysical Journal Letters 759, L40
18. McDonough WF, and Sun S-s (1995) The composition of the Earth. Chemical Geology 120:223–253
19. Morbidelli A, Bitsch B, Crida A, et al. (2016) Fossilized condensation lines in the Solar System protoplanetary disk. Icarus 267:368-376
20. Nittler LR, Chabot NL, Grove, TL, Peplowski PN (2018) The chemical composition of Mercury. In: Solomon SC, Nittler LR, Anderson BJ (eds) Mercury: The View After MESSENGER, Cambridge University Press, pp 30-51

21. Otegi JF, Bouchy F, Helled R (2020) Revisited mass-radius relations for exoplanets below 120 M_Earth. Astronomy & Astrophysics, 634:A43-55
22. Putirka KD, Rarick JC (2019) The composition and mineralogy of rocky exoplanets: a survey of >4,000 starts from the Hypatia Catalog. American Mineralogist 104:817-829
23. Schaefer L, Fegley B Jr (2017) Redox states of initial atmospheres outgassed on rocky planets and planetesimals. The Astrophysical Journal 843:120
24. Stewart AJ, Schmidt MW (2007) Sulfur and phosphorus in the Earth's core. Geophysical Research Letters 34:L131201
25. Thiabaud A, Marboeuf, U, Alibert Y, Leya I, et al. (2015) Elemental ratios in stars vs planets. Astronomy & Astrophysics 580:A30-37
26. Unterborn CT, Dismukes EE, Panero WR (2016) Scaling the Earth: A Sensitivity Analysis of Terrestrial Exoplanetary Interior Models. The Astrophysical Journal 819:32-40
27. Unterborn CT, Desch SJ, Hinkel NR and Lorenzo A (2018) Inward migration of the TRAPPIST-1 planets as inferred from their water-rich compositions. Nature Astronomy 2:297-302
28. Unterborn CT, Johnson JA, Panero WR (2015) Thorium abundances in solar twins and analogs: implications for the habitability of extrasolar planetary systems. The Astrophysical Journal 806:139
29. Weller MB, Lenardic A (2018) On the evolution of terrestrial planets: bi-stability, stochastic effects, and the non-uniqueness of tectonic states. Geoscience Frontiers 9:91-102
30. Yoshizaki T, McDonough WF (2020) The composition of Mars. Geochimica et Cosmochimica Acta 273:137-162